\documentclass{mn2e}
\usepackage{psfig}
\usepackage{mnras_cite}
\begin{document}

\def \vjec{\vfill\eject}
\def\kmsmpc{\mathrm{km}\ \mathrm{s^{-1}} \mathrm{Mpc^{-1}}}
\def\hmpc{h^{-1} \mathrm{Mpc}}
\def\hkpc{h^{-1} \mathrm{kpc}}
\def\msun{M_\odot}
\def\hmsun{h^{-1} M_\odot}
 \def\chisq{$\chi^2$}
\def \m3{{\rm Mark III}}
\def \etal {{et al.\ }}
\def \cf {{\it cf.\ }}
\def \vs {{\it vs.\ }}
\def \via {{\it via\ }}
\def \ie {{\it i.e.\ } }
\def \eg{{\it e.g.\ }}
\def\kms{\mathrm{km~s^{-1}}}
\def\br{{\bf r}}
\def\bv{{\bf v}}
\def\bV{{\bf V}}

\def\plotfour#1#2#3#4{\centering \leavevmode
\epsfxsize=.24\columnwidth \epsfbox{#1} \hfil
\epsfxsize=.24\columnwidth \epsfbox{#2} \hfil
\epsfxsize=.24\columnwidth \epsfbox{#3} \hfil
\epsfxsize=.24\columnwidth \epsfbox{#4}}

\def\gsim{~\rlap{$>$}{\lower 1.0ex\hbox{$\sim$}}}
\def\lsim{~\rlap{$<$}{\lower 1.0ex\hbox{$\sim$}}}
\def\d{{\rm d}}
\def\c{{\rm c}}

\newcommand{\pa}{\partial}
\newcommand{\cri}{_{\rm cr}}

\title[Dynamical-Friction Galaxy-Gas Coupling and Cluster
Cooling Flows]{Dynamical-Friction Galaxy-Gas Coupling and
Cluster Cooling Flows}

\author[A. A.~El-Zant, W.-T. Kim and M. Kamionkowski]
{Amr A. El-Zant$^1$, Woong-Tae Kim$^{2,3}$ \& Marc
Kamionkowski$^1$\\
$^1$Mail Code 130-33, California Institute of Technology,
Pasadena, CA 91125, USA\\
$^2$Harvard-Smithsonian Center for Astrophysics, 60 Garden Street, 
Cambridge, MA 02138, USA\\
$^3$Astronomy Program, SEES, Seoul National University, Seoul 151-742, Korea}

\maketitle

\begin{abstract}
We revisit the notion that galaxy motions can efficiently
heat intergalactic gas in the central regions of clusters through dynamical
friction.  For plausible values of the galaxy
mass-to-light ratio, the heating rate is comparable to the
cooling rate due to X-ray emission.  Heating occurs only for
supersonic galaxy motions, so the mechanism is 
self-regulating: it becomes efficient only when
the gas sound speed is smaller than the galaxy velocity dispersion.
We illustrate with the Perseus cluster, assuming a
stellar mass-to-light ratio for galaxies in the very central
region with the dark-matter contribution becoming comparable to
this at some radius $r_s$.  For $r_s \la 400~{\rm 
kpc} \sim 3 r_{\rm cool}$---corresponding to an average
mass-to-light ratio of $\sim10$ inside that radius---the
dynamical-friction coupling is strong enough to provide the
required rate of gas heating.  The measured sound speed is
smaller than the galaxy velocity dispersion, as required by this
mechanism.  With this smaller gas temperature and
the observed distribution of galaxies and gas, the energy
reservoir in galactic motions is sufficient to sustain the
required heating rate for the lifetime of the cluster.  The
galaxies also lose a smaller amount of energy through dynamical
friction to the dark matter implying that non--cooling-flow clusters
should have flat-cored dark-matter density distributions.
\end{abstract}



\section{Introduction}
\label{sec:intro}

Galaxy cluster gas loses thermal energy copiously through X-ray emission.
In the absence of energy input, radiative cooling in cores of clusters 
should result in  substantial gas inflow  (see Fabian 1994 for a review). 
These ``cooling flows'' would have associated mass deposition rates of 
several hundred ${\rm M_\odot yr^{-1}}$ in some clusters (Peres \etal 1998). 
Nevertheless, recent high-resolution X-ray observations 
(e.g., Peterson \etal 2001, 2003) have revealed that there is little 
evidence for the expected multi-phase gaseous structures, strongly suggesting 
that mass dropout is being prevented by some source that is heating the gas, 
thereby balancing radiative energy loss in  the central region of clusters.
Recent work  has focused on two prospective heating mechanisms: (1) 
diffusive heat transport, via thermal conduction 
(\pcite{tuc83,bre88,nar01,voi02,fab02,zak03,kim03a}) and/or turbulent mixing
(\pcite{cho03,kim03b,voi04}), from hotter gas in the outer region to that in
the core; (2) energy input from jets, outflows, and radiation from a central 
active galactic nucleus (\pcite{cio01,chu02,bru02,kai03}).

There is however at least one additional mechanism that appears to have been 
overlooked. It involves the energy lost by  concentrated clumps of matter 
(galaxies) as they move through the cluster. Part of this energy may go 
into re-arranging the dark-matter mass distribution  (El-Zant \etal 2003), 
but a  significant fraction  should end up deposited in the gas.
That dynamical-friction (DF) coupling can  transform the
dynamical energy of  galaxies into thermal motion in the gas has been known 
for at least four decades (e.g., Dokuchaev 1964).
Relatively recent work involving this notion includes the analysis by 
Miller (1986) of the Perseus cluster and that of Just \etal (1990) concerning the 
Coma cluster.  Both studies confirm that, provided that the mass-to-light ratio of 
galaxies is $\sim 20$, energy lost by galaxies to the gas  
should be sufficient in counteracting the radiative cooling of the gas in the central
region of these clusters.

Several developments on the empirical side suggest renewed relevancy for this 
mechanism. One obvious one involves  recent X-ray data confirming that a  
heating
mechanism {\em is} required; whereas the consensus in the 1980's was against 
this conclusion, it now seems inescapable. The second involves revisions
to the inferred gas electron densities
in the central region of clusters; best values referred to by Miller
and Just \etal are of the order of $10^{-3} {\rm cm^{-3}}$, while current 
best estimates are rather in the range of $10^{-2}-10^{-1} {\rm cm^{-3}}$ 
(Kaastra \etal 2004).  This leads to  an  order-of-magnitude increase in the 
dynamical friction coupling between the galaxies and gas. There has also been 
progress in determining the mass-to-light ratio of galaxies. On the theoretical 
side, work by Just \etal (1990) and Ostriker (1999) has since shed some light 
on the behavior of the dynamical-friction coupling in a gaseous medium in the 
transonic and subsonic regimes.

There appears to be {\em a priori} no reason why the rate of energy loss
from galaxies to gas   
via dynamical friction should be of the same order of that radiated by
the gas.  However, {\em this coupling is active only when
the sound speed of the gas is smaller than the typical velocity
of galaxies}.  The mechanism is therefore self-regulating; the
gas cools until the dynamical-friction heating rate is always
equal to the cooling rate.  We start by pointing out, in the
next Section, why this is expected to be so, before moving on to
develop a Monte-Carlo model to estimate the  total energy
expected to flow into the cooling region of the Perseus cluster,
using recent data for galaxy luminosities and projected
positions, as well as for the gas parameters of that cluster,
and averaging over a set of different realizations where
three-dimensional positions, velocities, and mass-to-light
ratios are treated as stochastic quantities.  The final Section
discusses briefly the central dark-matter distribution and
energetics, as well as some remaining questions.

\section{Dynamical friction in gaseous systems}
\label{sec:DF}

The fundamental difference between the dynamics of collisionless and 
gaseous systems is  the existence in the latter case of pressure
gradients which communicate forces at the sound speed $c_s$. If the 
galaxy velocity  $V$ is much larger than $c_s$ then
 the gravitational interaction between a particle at impact parameter $b$
leading to dynamical friction 
is unaffected by pressure forces (because by the time these are 
communicated to that region the galaxy is already at a distance 
$b \sqrt{1+(V/c_s)^2} \gg b$; e.g., Ruderman \& Spiegel 1971).
 If, on the other hand, $V < c_s$, pressure 
forces can be communicated to the perturbed region {\em before}
the minimum distance $b$ is achieved. 
The resulting pressure gradients ensure 
that the displacement 
of gas particles therein is hindered, 
and so the cumulative back reaction on the 
perturber that results in the dynamical-friction force is drastically
reduced.

In the collisionless case, when one speaks of a particle, one has 
in mind an actual material constituent. 
In the classic Chandrasekhar formulation, each of these
 contributes individually. The sum of the perturbations from particles moving faster
than the perturber (a galaxy) 
is negligible for small particle mass (being proportional to that mass). 
That from slower particles however results in a net contribution that
for the massive perturber  is independent of the smaller
background-particle mass, and is always directed opposite to the
perturber's motion. This results in the gradual decrease in the
dynamical-friction force when the perturber velocity falls below
that of the typical background particle. 
In the gaseous case however, unless the mean free
path is larger than the impact parameter, collisions will lead to rapid 
decoherence of individual particle motion. A coherent effect that results in 
dynamical friction should  then involve interactions between the perturber and 
gaseous elements instead of individual particles. 
If the bulk velocity, that is the macroscopic motion, of the gas 
elements is smaller than that of the perturber then all gas elements will contribute.

The above considerations suggest  that  gaseous dynamical friction will
follow the high-velocity approximation in a collisionless medium when $V > c_s$ and then drop 
sharply when $V < c_s$. This conclusion was already reached, on
the basis of steady-state perturbation theory (Ruderman \&
Spiegel 1971; Raphaeli \& Salpeter 1980)
and was confirmed by two more sophisticated, but rather 
different,  techniques  invoking
time dependence (Ostriker 1999) and fluctuation theory (Just \etal 1990).
Simulations by Sanchez-Salcedo \& Brandenburg (1999) 
have shown  broad agreement for particles moving on rectilinear 
trajectories and, more importantly, qualitatively confirmed 
the above conclusions in spherical systems (Sanchez-Salcedo \& Brandenburg 2001). 
For a perturber of mass $M_p$, moving with $V > c_s$ 
we will thus assume that energy is lost at a rate   given by 
\begin{equation}
\frac{d E}{d t} = \frac{4 \pi (G M_p)^{2} \rho_g}{V} \ln \frac{ b_{\rm max} }{ b_{\rm min} },
\end{equation}
and that the energy loss vanishes for $V < c_s$. 

The relevant density $\rho_g$ is the typical gas  density to be found within the cluster 
cooling radius
$r_{\rm cool}$, inside which the cooling time is less than a
Hubble time---that is, the region where heating must be invoked. 
If the galaxy 
is inside this radius,  the range of impact parameters that contribute to heating 
should, to a first approximation, correspond to a minimal scale determined by the galaxy size 
and a maximal one 
determined by the cooling radius. For these galaxies therefore we 
 take the Coulomb logarithm to be
$\ln \left(r_{\rm cool}/ r_0\right)$, where the ``galaxy radius'' $r_0$ is set to 10 kpc---about twice 
the effective radius and disk scale length for ellipticals and spirals,
respectively. Galaxies outside $r_{\rm cool}$ should also
contribute, but in this case the  contribution
from material inside $r_{\rm cool}$ to the energy lost by a galaxy
at radius $r$ comes from impact parameters  between 
${\rm max} (r_0, r - r_{\rm cool})$ and $r + r_{\rm cool}$.
This contribution is also restricted from within an  angle 
$2  \tan^{-1} [r_{\rm cool}/(r - r_{\rm cool})]$.
Replacing the minimal impact parameter with
$r- r_{\rm cool}$ leads to an error of $ < 1 \%$
in the results.
For galaxies outside $r_{\rm cool}$ therefore
\begin{equation}
\ln \frac{b_{\rm max}}{b_{\rm min}} \rightarrow \frac{1}{\pi}  \tan^{-1} \frac{r_{\rm cool}}{r - r_{\rm cool}}
\ln  \frac{ r + r_{\rm cool} } { r - r_{\rm cool} }.
\label{eq:barameters}
\end{equation}
For galaxies on highly eccentric trajectories, the bulk of the energy 
exchange 
with gas inside the cooling radius  takes place 
at closest approach  to  $r_{\rm cool}$.
However, corrections due to  this 
effect are not large, even if most galaxies are indeed on nearly
radial orbits. For any individual realization of such a highly anisotropic
 quasi--steady-state system, the dominant contribution to the
 energy input  will come from those galaxies that are at small radii
(since
both terms on the right hand side of~(\ref{eq:barameters})
rapidly decrease when $r$ is appreciably larger than $r_{\rm cool}$)
and are thus on their closest approach.

\section{The Perseus Cluster}
\label{sec:results}
Because of  the rather large energy input rate ($\sim 10^{45}~{\rm erg/s}$)
that is required to prevent a cooling flow,
the Perseus cluster is one of the more demanding cases for 
any cluster-heating model. In this Section 
we examine whether, even in this case, energy dissipated from 
cluster galaxies can be sufficient to provide the required 
power.

\begin{figure}
\psfig{file=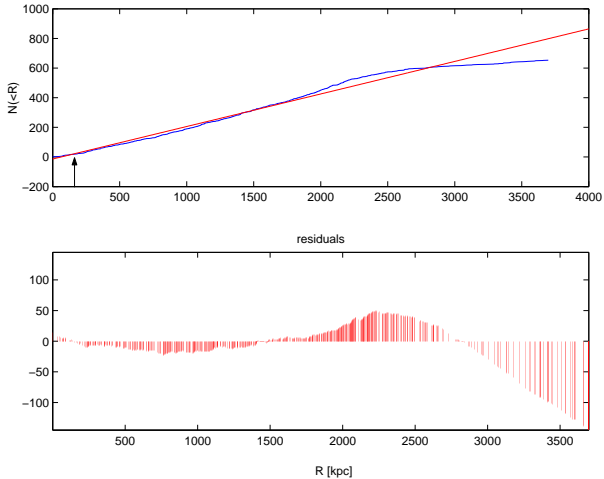,width=80mm}
\caption{
Number of galaxies inside  projected radius $R$,
which can be approximated by a linear fit (top) over
a large range  in radii 
and with limited relative error (bottom). The arrow indicates the cooling  
radius.}
\label{fig:MASS}
\end{figure}

Brunzendorf \&  Meusinger (1999)
have collected a catalog of projected positions 
and apparent luminosities of 660 galaxies in the Perseus cluster. 
In Fig.~\ref{fig:MASS} we plot the cumulative number of galaxies  as a 
function of projected radius $R$. 
As Fig.~\ref{fig:MASS} shows, this can be fitted over a large range in 
radius by a 
distribution $N(R) \sim R$, which implies a projected density distribution
$n(R) \sim 1/R$. This in turn implies a three dimensional density varying as $\sim 1/r^2$.
In this case, along the line of sight, and  at any given $R$, the number density 
of galaxies decreases as $\sim (R^2+Z^2)^{-1}$ (with $r^2 =R^2+Z^2$). 
This simple dependence of the number-density  distribution on
$Z$ is consistent with the latter being distributed in such  a 
way that $|Z| = R \tan^{-1} X$, with $X$ being a random variable uniformly 
distributed between 
0 and $\pi/2$. This is how the third spatial coordinate is chosen in
the random realizations of the Perseus cluster that we describe below. 
Since the density does drop more sharply than $1/r^2$ at large radii, this prescription will
slightly overestimate the $Z$ values of coordinates. This is however a minor
effect, since that region does not contain many galaxies and since most 
of the DF comes from inner cluster galaxies
(and if anything leads to a decrease in the total DF power pumped into 
$r_{\rm cool}$).

Galactic extinction is particularly severe at the position of the Perseus cluster in the sky.
Trials with the NED calculator show that corrections in the central region vary between 
$\sim -0.65$ and $\sim -0.9$, with a value of about $-0.7$ magnitude at the CD galaxy NGC 1275
(which is not included in the DF calculations).           
To all galaxies in the catalog we will apply an extinction correction  of  $-0.75$ magnitude. 
To this one needs to add a  
K-correction that ranges from -0.092 for ellipticals to -0.02 for late type spirals. 
The apparent
magnitude of disks are  however further affected by inclination effects. 
These vary as
$A_i \approx - \log (a/b)$, with $b/a \approx \cos i$  (assuming the axis ratio of the galaxy 
viewed edge on to be vanishingly small).  
This gives an average  correction $A_i = -0.19$. We also apply a 
cosmological dimming 
correction of 
$10 \log(1 + z) = -0.085$ to all galaxies.
Absolute  magnitudes are then obtained by assuming a distance to the Perseus 
cluster of $78~{\rm Mpc}$, compatible with a Hubble parameter
$H_0=100\,h$~Mpc with $h=0.7$.

Unless the structural parameters defining the density distribution 
 of elliptical galaxies vary significantly with mass, the 
Faber-Jackson relation and the virial theorem imply a luminosity-dependent
central mass-to-light ratio: $M/L \sim L^\gamma$, with 
$\gamma \sim 0.3-0.4$. Gerhard \etal (2001) have dynamically examined  the 
central mass-to-light ratio of a sample of elliptical galaxies and determined
that their results are in agreement with the assumption of structural homology, 
and that the variation of the mass-to-light ratio within an effective radius is 
largely due to change in stellar population. Although Trujillo \etal (2004) have argued 
that non-homology effects may be important in determining $\gamma$ (which they deduce 
to be $\sim 0.1$), a stellar mass-to-light variation with
$\gamma \sim 0.4$ also appears to exist for disk galaxies
(Salucci, Ashman \& Persic 1991). All these relations  exhibit significant scatter. 
We adopt a  zero-point mass-to-light ratio (which we will 
refer to as ``stellar'', even though the aforementioned papers do not 
exclude a modest dark-matter contribution in the inner regions)
consistent with these
results and containing a stochastic component to represent the scatter.
 Thus for a galaxy of luminosity $L$ we adopt
\begin{equation}
\left(\frac{M}{L}\right)_S = X \left(10 \frac{L}{5 \times 10^{10} \rm L_\odot}\right)^{0.3},
\label{eq:MLS}
\end{equation}
where $X$ is a random variable chosen from a normal distribution with 
average of unity  and dispersion $0.7$. (The normalization is chosen by 
inspection of Fig.~13 of Gerhard \etal 2001.)

The mass-to-light ratio is expected to increase with radius,
 as the dark-matter contribution becomes progressively 
more important (e.g., Gerhard \etal 2001; Takamiya  \&  Sofue 1999). 
For galaxies confined in clusters, the 
 contribution of dark matter at larger radii will depend on 
how much of the halo that should be surrounding the galaxy has
been removed. This will in turn depend  on the maximum excursion 
of that galaxy into the cluster center, and the density distributions 
of both the galaxy and the cluster.
For example, a singular  isothermal sphere with characteristic velocity dispersion $\sigma$ 
moving through
another isothermal system with dispersion $\Sigma > \sigma$, will
be cut to a radius $r_t = (\sigma/\Sigma) r_{\rm min}$ if $r_{\rm min}$
is the minimum distance that the centers of the two spheres achieve.
And so, in this approximation
of the situation,  an average  galaxy halo 
(velocity dispersion $\sim 200\,\kms$) moving through the Perseus cluster
(velocity dispersion $\sim 2000\,\kms$), will 
have  $r_t \sim  r_{\rm min}/10$ and the 
 mass  enclosed within this radius will be
$(2 \sigma^2/G) r_t$.
Since halos found in cosmological simulations have density 
profiles that are close
to singular isothermal spheres over a large range in radii, this representation
should be sufficient in approximating the initial conditions before any modification 
of these profiles takes place---provided that the density of the
galactic halos is scaled in such a way that the mass in the
very inner region is not dominated by the dark matter, so as to
be in line with the aforementioned studies concerning the
central mass-to-light ratios of galaxies.

\begin{figure}
\psfig{file=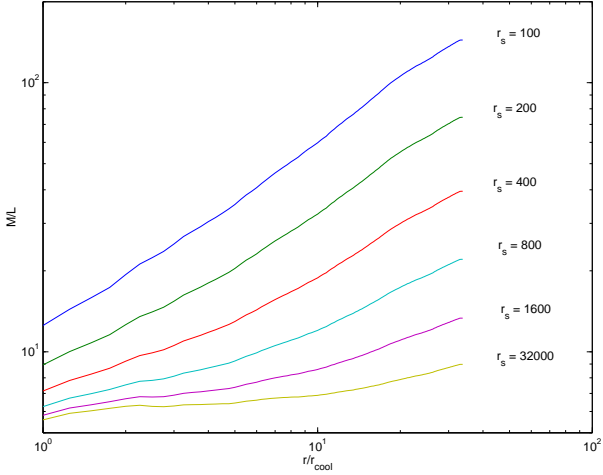,width=80mm}
\caption{
Mass-to-light ratios for galaxies
inside radius $r$ deduced from  
equation~\ref{eq:ML}
over 10,000 random  realizations of the Perseus cluster data.}
\label{fig:MLR}
\end{figure}

We will assume that the radial variation of the mass-to-light 
ratio of cluster galaxies is given by
\begin{equation}\label{eq:ML}
\frac{M}{L} = \left(1 + Y  \frac{r}{r_s}\right) \left(\frac{M}{L}\right)_S,
\end{equation}
where $(M/L)_S$ is defined by equation (\ref{eq:MLS}) and 
$r_s =  G M_{\rm lum}/\sigma^2 \sim  f (r/10)$ 
(with $f = r_{\rm min}/r$ and $M_{\rm lum}$ the luminous mass) defining
a spatial scale over which enough dark matter remains tied to
the galaxy for its contribution to the total mass to be
comparable to the luminous one.  Statistically, $f$ will  be 
determined by  the velocity distribution of galaxies,
with more anisotropic dispersions implying deeper entries into
the cluster core and so more stripping.
Uncertainty in this parameter are represented by an additional 
appeal to another random variable $Y$ again chosen from a normal
distribution with mean unity and dispersion 0.7.  In
Fig.~\ref{fig:MLR} we show the average mass-to-light ratio
within a given radius for the Perseus cluster averaged over
$10,000$ realizations of the random variables $X$ and $Y$.

\begin{figure}
\psfig{file=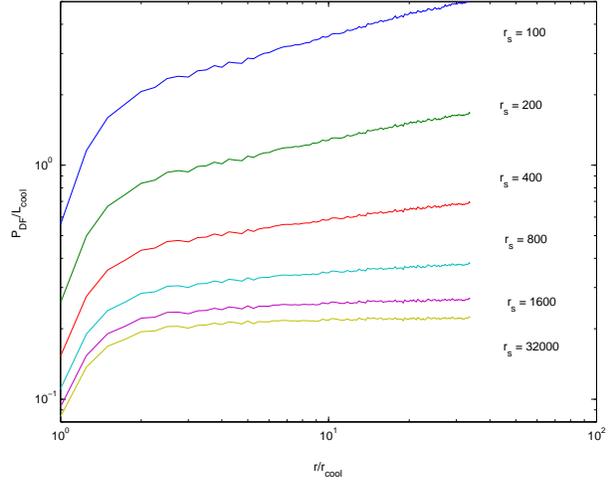,width=80mm}
\caption{
Power expected (over ten thousand random 
realizations of the cluster) to be transmitted 
to gas inside $r_{\rm cool}$ via dynamical friction 
on cluster galaxies within radius $r$, relative to the total 
power radiated from within $r_{\rm cool}$.}
\label{fig:DFE}
\end{figure}

Another dynamical variable determining the gas-galaxies  DF coupling is the 
velocity distribution of the latter.
Brunzendorf \& Meusinger have only determined the 
line-of-sight speeds of 169 of the galaxies in their samples. 
The velocity dispersions of galaxies, when binned radially in 
groups of  20 to 30, are $\sim 1000 - 1400$
$\kms$ with no obvious systematic variation incompatible with
small-sample statistics.
 Fig.~\ref{fig:DFE} shows the energy expected to be deposited via
dynamical friction into  $r_{\rm cool}$, from galaxies enclosed within a radius 
$r$, relative to the energy radiated from the cooling region,
for different values of the radial scale $r_s$. 
For each velocity component a normal distribution with zero mean and 
dispersion $\sigma$, such that $\sqrt{3} \sigma = 2000~\kms$, is assumed.
Results are  shown 
for the  10,000 realizations of the cluster in Fig.~\ref{fig:DFE},
assuming $r_{\rm cool} = 130~{\rm kpc}$, and the total energy emitted from 
$r < r_{\rm cool}$  to be $5 \times 10^{44} {\rm erg/s}$, consistent with the 
findings of Kaastra \etal (2004), when converted to correspond to 
$h = 0.7$. The sound speed $c_s = \sqrt{\gamma R T/ \mu}$ can also be read off
Table~5 of Kaastra \etal (2004).  Excluding the very inner data
point, which is probably influenced by  central activity
(Churazov \etal 2003), the central sound speed
starts at about $640~\kms$, eventually rising with radius to reach an asymptotic 
limit almost double that value, but remaining largely constant within most
of the cooling region (Churazov \etal Fig.~8). These studies also 
show  the  electron density to be slowly  varying 
within $r_{\rm cool}$, with values in the range $0.1 - 0.005~{\rm cm^{-3}}$
(there is also a factor of $\sqrt{0.7/0.5}$ that should be
included because we
use $h=0.7$, as opposed to the value $h=0.5$ used by these authors). 
In calculations shown  in Fig.~\ref{fig:DFE}, a value of $0.02~{\rm cm^{-3}}$ 
(setting  $\mu=1.18$ to convert electron densities into gas
density) and  $c_s = 700\,\kms$ are used.

\begin{figure}
\psfig{file=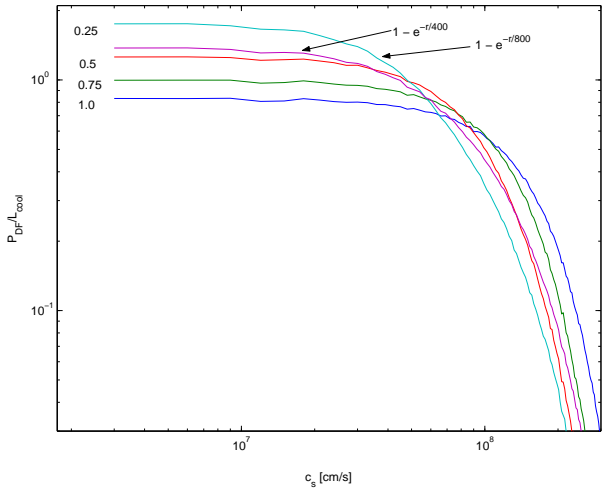,width=80mm}
\caption{
Same ordinate as in  Fig.~\ref{fig:DFE} but plotted against 
the gas sound speed for different (constant as well as radially varying) 
values of the ratio of the velocity
dispersions $\sigma_\theta/\sigma_r$ 
(and assuming $\sigma_\phi = \sigma_\theta$).}
\label{fig:ANIVEL}
\end{figure}

Because of the sharp cutoff in the DF-mediated  coupling when $V
< c_s$, the amount of  energy deposited into the gas must
sensitively depend on the relative magnitude of the galaxy
velocities and the sound speed when $V \sim c_s$. The former is
a function of the anisotropy of the distribution. For example,
if galaxies move on nearly  radial trajectories, their velocity
dispersion would be  closer to the line-of-sight value of 
1200 $\kms$, rather than 2000 $\kms$ as assumed above. The
simulations of Ghigna \etal (2000) suggest
that galaxies (as represented by cluster subhalos) may have
biased velocity dispersion in the 
central region, perhaps reflecting enhanced radial motion;
whereas at larger radii they have
velocity distributions that rather closely follow that of the dark matter. 
Fig.~\ref{fig:ANIVEL} shows the  power pumped into the gas relative to that
radiated within the cooling radius of the Perseus cluster, as predicted by our model, 
as a function of the sound speed for a range  of fixed 
and (radially) varying $\sigma_\theta/\sigma_r$ values (assuming $\sigma_\phi = \sigma_\theta$)
for $r_s = 400~{\rm kpc}$. It is rather remarkable that the competing effects of increased 
coupling at low velocities (because of the $1/V$ dependence), and the rapidly decreasing 
number of galaxies contributing to the DF  with increasing $c_s$, conspire to always 
ensure that a maximum value for the energy input is approached when 
$600\;\kms \la c_s \la 700\,\kms$, precisely the value 
attained in the (central) region where the temperature drops
in the cluster.

\section{Concluding remarks}
\label{sec:disc}

That dynamical-friction coupling between cluster galaxies and gas 
can significantly heat the latter component is not a new concept. Indeed 
Miller (1986) has pointed out that a few luminous galaxies  in the Perseus 
cluster's central  region may alone  deposit sufficient energy to 
keep gas in that region from cooling, provided that they had a mass-to-light
 ratio of about 20. We revisited this cluster using  
recently compiled galaxy and gas data 
and  determinations of  the mass-to-light ratios in galaxies.
Our Monte Carlo model  shows that the rate at which energy is 
deposited by galaxies into the gas within the cooling radius 
can completely compensate for radiative loss from within that 
radius, if galaxies at radii  $\la 400$ kpc
from the center of the cluster have dark mass that is comparable 
to their stellar mass (with galaxies at smaller radii having 
progressively smaller dark-matter content, vanishing for those near 
the center of the cluster).
The   associated  average  
mass-to-light ratio within this radius is about 10.

A robust and potentially important prediction of the model is that 
the drop in gas temperature invariably  observed in the cooling 
regions of clusters  is necessary 
for the DF coupling to be sufficiently strong. There is a natural
interpretation for this phenomenon: the gas cools
until dynamical-friction heating becomes sufficiently efficient
to prevent a
further drop in temperature.  The gas random motion is then 
coupled to that of the galaxies,  and so 
 remains in approximate equilibrium with it.
All this  simply follows 
from the fact that dynamical friction drops sharply for subsonic motion.
In the case of the Perseus cluster, the transition from weak to 
strong coupling  occurs precisely in the range of sound speeds
found in the cooling core of the cluster.

The action of the mechanism discussed in this paper also has consequences 
for the dark-matter distribution. As shown by El-Zant \etal (2003), dynamical 
friction from the galaxies will heat a   density cusp, creating 
a core  with radius corresponding to roughly a fifth 
of the original Navarro, Frenk \& White (1997) scale length 
(cf., Fig.~1 of El-Zant et al.), where dark matter has been spread out 
to larger radii (Fig.~3 of El-Zant et al.). 
Approximating the initial density
distribution in this
region such that  $\rho_i = \rho_0 (r_0/r)$ and the final one
with $\rho_f = \rho_0$, the energy required for this transformation 
is $\Delta \Phi = 
(8 \pi G)^{-1}  \int_{0}^{r_0} 
(|d \phi/dr|_i^2 - |d\phi/dr|_f^2) 4 \pi r^2 dr =
22 \pi^2 \rho_0^2 r_0^5/45$ [with $d\phi/dr
= (4 \pi G/r^2 )\int \rho r^2 dr$].
Taking $r_0 = r_{\rm cool}$ and 
$\rho_0$ to correspond to four times the value defined
by the electron density $\rho_e (r_0) = 0.0053$ (where we have
used the gas to  gravitational mass ratio
in Table~5 of Peres \etal 1998 and equation (4) of Churazov \etal 2003 
for the electron density), 
one finds $\Delta \Phi = - 6 \times 10^{60}~{\rm erg}$, which is comparable to the
binding energy of the CD galaxy $\Phi_{\rm CD} \approx - 
(G M_{\rm CD}^2/R_{\rm CD})$.  
It is however significantly smaller than the energy radiated 
from gas inside the cooling radius for the age of the cluster (e.g., for five Gyr this 
amounts to  $8 \times 10^{61}\, {\rm erg}$). Nevertheless, this energy is  easily available from fast 
moving galaxies from beyond the very inner region---for example its value
 coincides with the  kinetic energy in a mass similar to that of the CD 
galaxy and moving at $2000~\kms$, and that material can be supplied  
solely by the stellar mass of galaxies within $2 r_{\rm cool}$---the 
region from which, according to our model, the bulk of energy input to the gas 
comes from.

It is worth noting here that, under the circumstances just described, the mass of the gas 
within $r_{\rm cool}$ [which using equation (4) of Churazov \etal 2003,
adjusted for $h=0.7$
 with $r_{\rm cool} = 130\,{\rm kpc}$ amounts to $2.2 \times
 10^{12} {\rm M_\odot}$] is comparable 
to that of the galaxies in that region. Due to the temperature drop in the gas, thermal motions
can be significantly smaller than that of the galaxies. The energy of the gas can therefore be
smaller than that of the galaxies by up to 
an order of magnitude. There is therefore sufficient energy in galaxies to heat the gas many 
times over,  with the crucial parameter, which was the focus of this paper,
 being the rate at which this is transferred. 
Furthermore, once the dark matter in the center has been heated, it absorbs little additional 
energy from the galaxies, since  dynamical friction does not act on fast particles, 
and the coupling with the dark matter at larger radii decreases rapidly with 
radius (see El-Zant \etal 2003 for further discussion).  The
energy lost to the dark matter is therefore less than that going
into  keeping the gas at constant temperature.
The mass of the CD galaxy, as well as the spatial and velocity 
distribution of galaxies in the central region of the cluster will
reflect the history of energy loss to both components.

Several questions remain open. Prominent among these is the 
issue of thermal stability. From the condition that thermal stability
requires a remarkably narrow range of heating rates,
Bregman \& David (1989) have 
argued against Miller's proposition that the Perseus cluster gas is heated
by DF from galaxies.  However the functional
form of the heating rate used by these authors does not agree with later
calculations that have been borne out by numerical studies---e.g., their 
postulated form has a {\em negative} energy transfer rate ($\sim -1/V^3$)
for highly subsonic velocities, resulting in gas {\em cooling}, which is 
 incompatible with the positive (even if small)  heating found 
by Just \etal (1990) and 
Ostriker (1999) in that limit. 
It will be therefore necessary to repeat these calculations using 
 adequate forms for the heat-loss function. These calculations would also 
address another crucial question, that concerning  the precise manner
in which  the energy lost by the galaxies is distributed in the gas.  
We have 
adopted the simple approach where this energy is deposited
isotropically and equally in logarithmic intervals. This 
should approximate how the energy is initially deposited, at least
for highly supersonic galaxies (and 
since, in the cooling region,  the sound speed may be several times
smaller than the  velocity dispersion, this may not be a very 
bad assumption). 
Even then, however, the deposited  energy  may still  be
transported toward the center of the cluster by wave motions, 
as suggested, for example,  by Balbus \& Soker (1990), and
redistributed.  We note here that the claim of these authors
that DF from galaxies is insufficient to heat the cluster core
is based upon outdated values for the electron density and an
outdated ($1/r$) gas distribution.  With updated values for the
electron density and gas distribution (e.g., the empirical
formula of Churazov \etal 2003), their conclusions are
changed, as we have shown.  We thus conclude that it is far from
obvious that DF-heating of cluster gas is irrelevant.

\section*{ACKNOWLEDGMENTS}
The authors would like to thank M. C. Begelman, E. Bertschinger,
P. Goldreich, R. Narayan, E. Ostriker, N. Scoville and R. Sunyaev for helpful 
discussions and communications.  This work was supported at Caltech by NASA
NAG5-9821, DoE DE-FG03-92-ER40701 and NSF grant  AST 00-98301,
and at the CfA by NAG5-10780 and NSF grant AST 0307433.


\begin{thebibliography}{}

\bibitem[Balbus \& Soker<1990>]{BalSok93} Balbus S. A., Soker
     N., 1990, ApJ, 357, 353  
\bibitem[Bregman \& David<1988>]{bre88}
   Bregman J.\ N., David L.\ P., 1988, ApJ, 326, 639

\bibitem[Bregman \& David<1989>]{BreDav89} Bregman J.\ E., David
     L.\ P., 1989, ApJ, 341, 49

\bibitem[Br\"uggen \& Kaiser<2002>]{bru02}
   Br\"uggen M., Kaiser C.\ R.,  2002, Nature, 418, 301


\bibitem[Brunzendorf \& Meusinger<1999>]{BruMeu99} Brunzendorf
     J., Meusinger H., 1999, A\&AS, 139, 141  

\bibitem[Cho et al.<2003>]{cho03}
   Cho J., Lazarian A., Honein A., Knaepen B., Kassinos S.,
   Moin S., 2003, ApJ, 589, L77

\bibitem[Churazov et al.<2002>]{chu02}
   Churazov E., Sunyaev R., Forman W., B\"ohringer H., 2002,
   MNRAS, 332, 729

\bibitem[Churazov et al <2003>]{Chura03} 
Churazov E., Forman W., Jones C., B\"ohringer H., 2003, ApJ, 590, 225

\bibitem[Ciotti \& Ostriker<2001>]{cio01}
   Ciotti L., Ostriker J.\ P., 2001, ApJ, 551, 131

\bibitem[Dokucahev<1964>]{Dok64} Dokuchaev V. P., 1964, Sov. Astron., 8, 23 

\bibitem[El-Zant et al.<2003>]{ElZetal93} El-Zant A. A.,
     Hoffman Y., Primack J. P., Combes F., Shlosman I.,
     astro-ph/0309412

\bibitem[Fabian<1994>]{Fab94} Fabian A. C., 1994, ARAA, 32, 277

\bibitem[Fabian, Voigt, \& Morris<2002>]{fab02}
   Fabian A.\ C., Voigt L.\ M., Morris R.\ G., 2002, MNRAS,
   335, L71


\bibitem[Gerhard \& Kronawitter<2001>]{GerKro01} Gerhard O.,
     Kronawitter A., 2001,  AJ, 121, 1936

\bibitem[Ghigna et al.<2000>]{Ghietal00} Ghigna S., Moore B.,
     Governato F., Lake G.,Quinn T., Stadel J., 2000, ApJ,
     544, 616

\bibitem[Just et al.<1990>]{Jusetal90} Just A., Deiss
     B. M., Kegel W. H., Bohringer H.,  Morgil G. E., 1990,
     ApJ, 400, 353 

\bibitem[Kaastra<2004>]{Kaaetal04} Kaastra J. S. \etal, 2004, A\&A,
     413, 415 
\bibitem[Kaiser \& Binney<2003>]{kai03}
   Kaiser C.\ R., Binney, J., 2003, MNRAS, 338, 837


\bibitem[Kim \& Narayan<2003a>]{kim03a}
     Kim W.-T., Narayan R., 2003, ApJ, 596, 889

\bibitem[Kim \& Narayan<2003b>]{kim03b}
     Kim W.-T., Narayan R., 2003, ApJ, 596, L139

\bibitem[Milleter<1986>]{Mil86} Miller L., 1986, MNRAS, 220, 713 

\bibitem[Narayan \& Medvedev<2001>]{nar01}
   Narayan R., Medvedev M.\ V., 2001, ApJ, 562, L129

\bibitem[Navarro, Frenk \& White<1997>]{Navetal97} Navarro
     J. F., Frenk C. S., White S. D. M., 1997, ApJ, 490, 493

\bibitem[Ostriker<1999>]{ost99}
   Ostriker E., 1999, ApJ, 513, 252

\bibitem[Peres et al.<1998>]{Peretal98} Peres C. B., Fabian
     A. C., Edge A. C., Allen S. W., Johnstone R. M., White
     D. A., 1998, MNRAS, 298, 416

\bibitem[Petetal.<2001>]{Petetal01} Peterson J. R. \etal, 2001,
     A\&A, 365, L104

\bibitem[Petetal.<2003>]{Petetal03} Peterson J. R. \etal, 2003,
     A\&A, 590, 207 

\bibitem[Raphaeli \& Salpeter<1980>]{RapSal80} Raphaeli Y.,
     Salpeter E. E., 1980, ApJ, 240, 20

\bibitem[Ruderman \& Spiegel<1971>]{RudSpi97} Ruderman M. A.,
     Spiegel E. A., 1971, ApJ, 151, 679

\bibitem[Salucci, Ashman \& Persic<1991>]{SalAshPer91}
Salucci P, Ashman K. M.,  Persic M., 1991,  ApJ, 379, 89

\bibitem[Sanchez-Salcedo \& Brandenburg<1999>]{SanBra99}
     Sanchez-Salcedo F. J., Brandenburg A., 1999, ApJ, 522, L35 


\bibitem[Sanchez-Salcedo \& Brandenburg<2001>]{SanBra01}
     Sanchez-Salcedo F. J., Brandenburg A., 2001, MNRAS, 322, 67 


\bibitem[Takamiya \& Sofue<2000>]{TakSof00} Takamiya T., Sofue
     Y.,  2000, ApJ, 534, 670

\bibitem[Trujillo, Burkert \& Bell<2004>]{TruBurBel04} Trujillo
     I., Burkert A., Bell E. F., 2004, ApJ, 600, L39

\bibitem[Tucker \& Rosner<1983>]{tuc83}
   Tucker W.\ H., Rosner, R., 1983, ApJ, 267, 547

\bibitem[Voigt et al.<2002>]{voi02}
   Voigt L.\ M., Schmidt R.\ W., Fabian A.\ C., Allen S.\ W.,
   Johnstone R.\ M.n 2002, MNRAS, 335, L7


\bibitem[Voigt \& Fabian<2004>]{voi04}
    Voigt L., Fabian A., 2004, MNRAS, 347, 1130

\bibitem[Zakamska \& Narayan<2003>]{zak03}
   Zakamska N.\ L., Narayan R., 2003, ApJ, 582, 162


\end{thebibliography}
\end{document}